\documentclass{Rinton-P9x6}\usepackage{amsmath,amssymb}
\def\<{\langle}\def\>{\rangle}
\def\Tr{\operatorname{Tr}}
\def\map#1{{\mathrm{#1}}}
\def\set#1{{\sf #1}}

\def\Bnd#1{\set{B(#1)}}
\def\sH{\set{H}}\def\sK{\set{K}}\def\sA{\set{A}}
\def\sF{\set{F}}\def\sR{\set{R}}\def\sP{\set{P}}

\def\eg{e. g. }\def\ie{i. e. }
\def\N#1{\left|\!\left|#1\right|\!\right|}\def\n#1{|\!|#1|\!|}
\catcode`@=11  
\def\@cite#1{$^{#1}$}
\def\@refe#1{[#1]}
\def\@biblabel#1{{\normalsize\bf{#1}}}
\def\refe{\@ifnextchar [{\@tempswatrue\@citexr}{\@tempswafalse\@citexr[]}}
\def\@citexr[#1]#2{\if@filesw\immediate\write\@auxout{\string\citation{#2}}\fi
  \def\@citea{}\@refe{\@for\@citeb:=#2\do
    {\@citea\def\@citea{,}\@ifundefined
       {b@\@citeb}{{\bf ?}\@warning
       {Citation `\@citeb' on page \thepage \space undefined}}%
\hbox{\csname b@\@citeb\endcsname}}}{#1}}
 \catcode`@=12

\title{The quantum bit commitment:
a complete classification of protocols}
\author{Giacomo Mauro D'Ariano}
\address{{\em Quantum Optics \& Information Group} of the INFM\\
Dipartimento di Fisica ``A. Volta'', via Bassi 6, I-27100 Pavia, Italy \medskip \\
Department of Electrical and Computer Engineering,\\ 
Northwestern University, Evanston, IL  60208
}
\begin{document}
\maketitle
\abstracts{This paper addresses the controversy between 
Mayers, Lo and Chau\cite{MayersLoChau} on one side, and 
Yuen\cite{yuenqbc} on the opposite side, on whether there exist or not 
unconditionally secure protocols. For such purpose, a 
complete classification of all possible bit commitment protocols is
given, including all possible cheating attacks. For the simplest class
of protocols (non-aborting and with complete and perfect
verification), it is shown how naturally a game-theoretical situation
arises. For these protocols, bounds for the cheating probabilities 
are derived, which turn out to be different from those given in the
impossibility proof\cite{MayersLoChau}. The whole classification
and analysis has been carried out using a {\em finite open system}
approach. The discrepancy with the impossibility proof is 
explained on the basis of the implicit adoption of a {\em closed
system approach}---equivalent to modeling the commitment as performed
by two fixed machines interacting unitarily in a overall {\em closed
system}. However, it is shown that the closed system approach for the
classification of commitment protocols unavoidably leads to 
infinite dimensions, which then invalidate the continuity argument at
the basis of the impossibility proof.}
\section{Introduction}
It is of practical relevance to establish if there exist secure
quantum bit commitment protocols, since the quantum bit commitment is
a crucial element to build up more  sophisticated protocols, such as
remote quantum gambling, coin tossing, and unconditionally  secure
two-party quantum computation.
\par In the bit commitment Alice provides Bob with a piece of evidence that
she has chosen a bit $b=0,1$ which she commits to him. Later, Alice
will open the commitment, revealing the bit $b$ to Bob, and proving
that it is indeed the committed bit with the evidence in Bob's possession.  
Therefore, Alice and Bob should agree on a protocol which satisfies
simultaneously the three requirements: (1) it must be {\em
concealing}, namely Bob should not be able to retrieve $b$ before the
opening;  (2) it  must be {\em binding}, namely Alice should not be
able to change $b$ after the commitment; (3) it must be
{\em verifiable}, namely Bob must be able to check $b$ against the
evidence in his possession, according to the rules of the 
protocol. In a in-principle proof of security of the commitment it is
supposed that both parties possess unlimited technology, \eg
computational power, space, time, etc., and the protocol is said   
{\em unconditionally secure} if neither Alice nor Bob can cheat with
significant probability of success as a consequence of physical laws.

In 1993, a quantum mechanical protocol was proposed\cite{BCJL}, and
the unconditional security of this protocol has been generally
accepted for long time. The insecurity of this protocol was shown by 
Mayers, Lo and Chau\cite{MayersLoChau} in 1997, where it was
recognized the possibility for Alice to cheat by entangling the
committed evidence with a quantum system in her possession, and it was
argued that no unconditionally secure protocol is possible. Finally after
2000 Yuen\cite{yuenqbc} presented some protocols
which challenged the previous impossibility proof, mostly on the basis
of the possibility of encoding the bit on an {\em anonymous state}
given to Alice by Bob and known only to him, and suggesting
the use of {\em decoy systems} that make the protocol concealing in
the limit of infinitely many systems, with the possibility for Bob of
performing his quantum measurement before Alice opening, whence
disputing the general availability of EPR cheating for Alice. 

In this paper, in order to provide clarifications on the controversy
we will present a classification of all possible bit commitment
protocols based on a single commitment step, analyzing the main
cheating strategies for both parties (a full derivation of the classification,
the reduction of multi-step commitments  to a
single step, and a more exhaustive analysis of cheating attacks can be
found in Ref.\refe{mine}, of which the present paper is a much shorter
version).  For the simplest class of protocols (non-aborting, with
complete and perfect verification) we will show 
how naturally a game-theoretical situation arises. Bounds for the
cheating probabilities of these protocols are presented, which are
different from those given in the impossibility
proof\cite{MayersLoChau}. 
In the final discussion we will see how the discrepancy between the
two opposite analysis arises, due to the implicit adoption in the
impossibility proof of a {\em closed system approach}, equivalent to
modeling the commitment as performed 
by two fixed machines interacting unitarily in a overall {\em closed
system}. However, it is shown that such modeling, along with the
requirement of unlimited technology, necessarily lead to infinite
dimensions,  which invalidate the continuity argument at the basis of
the impossibility proof.  
\section{The classification of protocols}
The most general bit commitment scheme with a single step is of the 
form: (1) Bob prepares the Hilbert space $\sH$ with the {\em anonymous
state} $|\varphi\>\in\sH$, and sends $\sH$ to Alice; (2) Alice {\em
modulates} the value $b=0,1$ of the committed bit on the anonymous state 
$|\varphi\>$ and sends the output back to Bob. The {\em bit
modulation} is a quantum operation (QO) $\map{M}^{(b)}$
parametrized by $b$. 
Such scheme contains all possibilities, including
Yuen's protocols\cite{yuenqbc}, and the protocols considered by
Mayers, Lo and Chau\cite{MayersLoChau}, which correspond 
to {\em openly known} $|\varphi\>$. In general the output Hilbert
space $\sK$  of the QO will be different from $\sH$, since Alice can 
send back to Bob a quantum system different from what he sent to her.
\par In Ref.\refe{mine} a complete classifications of all possible
protocols is derived, on the basis of the fact that since Alice has unlimited
technology,  she can always achieve the encoding QO's $\map{M}^{(b)}$
of the committed bit value $b$ via a {\em perfect pure
measurement}. For {\em non aborting protocols}, this corresponds to
the following QO's  
\begin{eqnarray}
\map{M}^{(b)}(|\varphi\>\<\varphi|)
=\Tr_{\sF\otimes\sP}[U^{(b)}(|\varphi\>\<\varphi|\otimes|\omega\>\<\omega|_\sA\otimes\rho_\sP)
U^{(b)}{}^\dag],\label{mmap}
\end{eqnarray}
where $\sA$ is the preparation ancilla/decoy Hilbert space prepared in the
state $|\omega\>$; $\sF$ is the measurement ancilla Hilbert space
on which Alice performs a complete von Neumann measurement, and we
have that $\sK\otimes\sF\simeq\sH\otimes\sA$; 
$\sP$ is the space of the {\em secret parameter}, say $j$, which is
needed in order to make the protocol {\em concealing} and at the 
same time {\em verifiable} (so that the modulation is actually a
choice between two {\em ensembles} of QO's $\{\map{M}_j^{(b)}\}$ for
$b=0,1$). Therefore, the best option for Alice is to achieve the
encoding QO by preparing the ancilla/decoy state
$|\omega\>_\sA\in\sA$, performing the unitary transformation $U^{(b)}$
on $\sH\otimes\sA$, making a complete von Neumann measurement on
$\sF$, with outcome say $i$, and finally send $\sK$ to Bob. The
partial trace on $\sF\otimes\sP$ on the basis $\{|i\>\otimes|j\>\}$,
which describes Alice's measurement, corresponds to
the Kraus decompositions  $\map{M}^{(b)}=\sum_{ji}p_j E_{ji}^{(b)}\cdot
E_{ji}^{(b)}{}^\dag$, 
where $j$ is the {\em secret parameter} and $i$ is the {\em secret outcome}, 
and the probabilities $p_j=\< j|\rho_\sP|j\>$ for 
$j$ will depend on the preparation $\rho_\sP$. 
In a protocol which is completely and perfectly verifiable Alice  
tells $b$, $j$ and $i$ to Bob, who verifies the state 
$E_{ji}^{(b)}|\varphi\>$.
Since the local QO's on $\sK$ and $\sF\otimes\sP$ commute, 
Alice has the possibility of: (1) first sending $\sK$ to Bob; (2) then
performing the measurement on $\sF\otimes\sP$ at the very last moment
of the opening. As we will see, this is the basis for Alice EPR
cheating attacks. Notice that strictly trace-decreasing QO's---\ie
aborting protocols---pose limitations to Alice's EPR cheating. In
fact, Alice cannot delay the abortion of the protocol at the opening,
and must declare it at the commitment. Since both secret parameters
$j$ and $i$ can be conveniently measured by Alice, they can be treated
on equal footings as a single parameter $J\equiv(j,i)$.  With the notation
$E_J^{(b)}\doteq\sqrt{p_j}E_{ji}^{(b)}\in\Bnd{\sH,\sK}$, the maps
write
\begin{equation}
\map{M}^{(b)}(|\varphi\>\<\varphi|)=
\sum_j p_j\map{M}_j^{(b)}(|\varphi\>\<\varphi|)=
\sum_J E_J^{(b)} |\varphi\>\<\varphi|E_J^{(b)}{}^\dag,
\end{equation}
\section{Cheating}
For a discussion of all possibilities of cheating see Ref. \refe{mine}. 
Here we analyze the only the useful attacks by both Alice and Bob.
\smallskip
\par {\bf Alice cheating.} After the commitment and before the
opening Alice can try to cheat by performing a unitary transformation
$V$ on $\sF\otimes\sP$: this is the so-called EPR attack. 
Without changing the QO's $\map{M}^{(b)}$, the maneuver
will change their Kraus decompositions---which are relevant at the
opening---as $\{E_J^{(b)}\}\rightarrow\{E_J^{(b)}(V)\}$, keeping the
cardinality, in the following way  
\begin{equation}
E_J^{(b)}(V)=\sum_L E_L^{(b)} V_{JL},\qquad V_{JL}= \<J |V|L\>.
\end{equation}
\par The probability that Alice can cheat successfully in 
pretending having committed, say, $b=1$, whereas she 
committed $b=0$ instead, is given by
\begin{equation}
P_c^A(V,\varphi)=\sum_J 
\frac{|\<\varphi|E_J^{(0)}{}^\dag(V)
E_J^{(1)}|\varphi\>|^2}{\n{E_J^{(1)}\varphi}^2},\label{PcAV}
\end{equation}
and depends on the anonymous state $|\varphi\>$ and on
the cheating transformation $V$. Without any knowledge of
$|\varphi\>$, the best that Alice can do is to adopt a conservative
strategy, by maximizing her probability of cheating in 
the worst case, corresponding to the {\em minimax} choice of $V$
\begin{equation}
(P_c^A)_\mu\doteq\max_V\min_\varphi 
P_c^A(V,\varphi).\label{minimaxP}
\end{equation}
It is evident that in this way a game theoretical situation arises, in
which Bob choses $|\varphi\>$ and Alice choses $V$, with the 
probability $P(V,\varphi)$ playing the role of a {\em payoff 
matrix}. The actual game situation is more complicated---due
for example to Bob cheating---and will be analyzed elsewhere.
\smallskip\par{\bf Bob cheating.} 
Bob can try to cheat by making the {\em best
discrimination} between the two maps $\map{M}^{(b)}=\sum_j
p_j\map{M}_j^{(b)}$. However, since he doesn't know the probabilities 
$p_j$ actually used by Alice, his strategy will be suboptimal,
and his actual cheating probability $P_c^B$ will be lower than 
the probability $(P_c^B)_{opt}$ corresponding to the optimal 
strategy with the right probabilities $p_j$. Since map-discrimination
is generally more reliable with the map acting locally on an entangled 
state\cite{entang_meas}, instead of preparing $|\varphi\>\in\sH$ Bob  
prepares an entangled state on $\sH\otimes\sR$ and sends only $\sH$ to
Alice. Therefore, for equally probable bit values $b=0,1$, Bob's
optimal probability of cheating is bounded as follows\cite{mine}   
\begin{equation}
P_c^{B}\le(P_c^{B})_{opt}=\frac{1}{2}+\frac{1}{4}\N{\map{M}^{(1)}-\map{M}^{(0)}}_{cb},\label{cbcheat}
\end{equation}
where $\N{\cdot}_{cb}$ denotes the completely bounded (CB) 
norm. 
\smallskip\par{\bf Bounds for cheating probabilities.}
If the protocol is perfectly concealing the CB-norm in
Eq. (\ref{cbcheat}) is zero, and the two maps are the same, whence 
the their Kraus are connected via a unitary transformation $V$ on
$\sF\otimes\sP$, and Alice can cheat with probability one.
Let's consider now the case in which  Bob's optimal probability of
cheating $(P_c^{B})_{opt}$ is infinitesimally close to $\frac{1}{2}$, namely
$\n{\map{M}^{(1)}-\map{M}^{(0)}}_{cb}=\varepsilon$. Notice that
generally $\varepsilon$ is vanishing for increasing dimension of $\sK$
(such as when the approximately concealing condition is 
achieved for increasingly large number of decoy systems\cite{yuenqbc}),
and no obvious continuity argument can be invoked to assert that Alice
cheating probability will approach unit for vanishing
$\varepsilon$. More precisely, in the present context the continuity
argument of Ref. \refe{MayersLoChau} would imply that
\begin{equation}
1-(P_c^A)_\mu=\omega
\left(\N{\map{M}^{(1)}-\map{M}^{(0)}}_{cb}\right),
\quad \lim_{\varepsilon\to0}\omega(\varepsilon)=0
\label{impossibility}
\end{equation}
with the function $\omega(\varepsilon)$ independent on the 
dimension of $\sK$.
However, using anonymous states such assertion may 
turn out to be false. In fact, it is obvious that if there is an 
alternate Kraus 
decomposition $\{E_J^{(0)}(V)\}$ for the map $\map{M}^{(0)}$
such that the two Kraus $\{E_J^{(0)}(V)\}$ and $\{E_J^{(1)}\}$ are
close, then the protocol is approximately concealing and not 
binding, since\cite{mine}
\begin{eqnarray}
(P_c^{B})_{opt}-\frac{1}{2}&=& 
\frac{1}{4}\N{\map{M}^{(1)}-\map{M}^{(0)}}_{cb}\le
\frac{1}{2}\sqrt{\N{\sum_J\left|E_J^{(0)}(V)-E_J^{(1)}\right|^2}},
\label{boundbob}\\
P_c^A(V,\varphi)&\ge&\left[1-\frac{1}{2} 
\N{\sum_J\left|E_J^{(0)}(V)-E_J^{(1)}\right|^2}\right]^2,
\label{boundalic}
\end{eqnarray}
where for any operator $A$ we use the customary abbreviation 
$|A|^2\doteq A^\dag A$.
However, the impossibility proof would be true if a bound of the 
form (\ref{boundbob}) would be satisfied in the reverse direction,
in which case one would have
\begin{equation}
1-(P_c^A)_{\mu,av}\le \min_V
\N{\sum_J\left|E_J^{(0)}(V)-E_J^{(1)}\right|^2}\le
\omega\left(\N{\map{M}^{(1)}-\map{M}^{(0)}}_{cb}\right),
\label{endofstory}
\end{equation}
which would correspond to the following {\em continuity 
argument}: if two CP maps are close in CB-norm, then
for a given fixed Kraus decomposition for one of the two maps, 
there is always an alternate Kraus decomposition for the other 
map such that the two are close. 
Since as regards the cheating probabilities we have considered 
only the case of non-aborting protocols with perfect-verification, 
proving the continuity argument (\ref{endofstory}) or directly the
bound (\ref{impossibility}) would means that a secure protocol 
can still be searched outside such class of protocols. On the other 
hand, finding a counterexample to Eq. (\ref{impossibility})
would provide a perfectly verifiable and unconditionally secure 
protocol. 
\section{Discussion}
The discrepancy between the previous analysis and the analysis
beneath the impossibility proof\cite{MayersLoChau} 
is essentially due to the fact that the latter is based on the
assumption that the starting state of the commitment protocol
is openly known, in the sense that the probability distribution of
the state is given, and then the corresponding mixed state can be 
purified. The general underlying idea is that the protocol should
be processed by {\em machines}, and therefore all probability 
distributions are defined, and purified inside the machines. 
However, such an assumption is certainly not realistic for a 
cryptographic protocol, where each party has actually the 
freedom of changing or tuning the machine, namely chosing 
any desired probability distribution. 
One can continue to argue on this line, asserting that 
changing the machine is equivalent to use a larger machine. However, 
this will be equivalent to consider {\em infinite machines},
corresponding to uniform probabilities on infinite sets, and this
would invalidate an impossibility proof based on a non proved
continuity argument. 
\par The above hill-posed mathematical framework arises from the
Bayesian approach to secret parameters, dictated from the {\em closed
system} modeling with fixed machines and purification of
probabilities.  Alternative to the previous approach, we have the realistic
{\em finite open system} approach, in which unknown parameters are
treated as such, without the need of any a priori probability
distribution, in which we can address the problem for finite dimension
with the parameter $\varepsilon$ depending on it. Then, if one  
proceeds by treating unknown parameters as such, no openly 
known state can be assumed, and the anonymous state encoding 
of Yuen\cite{yuenqbc} leads to the present classification of
protocols. Notice that if the initial state $|\varphi\>$ is openly
known, then for that given fixed states all QO's can be regarded as
random unitary transformations (since all states are connected by
unitary transformations), and this lead to the simple form of Alice 
cheating probability in terms of fidelities\cite{MayersLoChau},
whereas in the present context the probability of cheating has the
more involved form (\ref{PcAV}), due to the fact that the state 
$|\varphi\>$ is unknown, and that there are QO's that don't admit 
random unitary Kraus decompositions.
\par Finally, regarding the possibility of aborting protocols, 
one could always reasonably adopt equivalent protocols which don't
abort, since the repeated commitment will eventually be successful.
However, such kind of protocols will necessarily be infinite  
convex combinations of protocols on infinite dimensional 
anonymous spaces $\sH$, and again one the {\em closed system}
approach would necessarily lead to infinite dimensions.
\bigskip\par{\bf Acknowledgments.}
This work has been jointly founded by the EC under the program 
ATESIT (Contract No. IST-2000-29681) and by the USA Army Research
Office under MURI Grant No. DAAD19-00-1-0177. Extensive discussions
with H. P. Yuen, who motivated this work, are  acknowledged.
 
\end{document}